\documentclass[aps,pre,preprint,superscriptaddress,showpacs]{revtex4-1}
\usepackage{graphicx}
\usepackage{amsmath}
\usepackage{amssymb}
\usepackage[usenames]{color}

\newcommand\st{\bgroup\markoverwith
{\textcolor{blue}{\rule[0.5ex]{2pt}{0.4pt}}}\ULon}

\begin{document}

\title{Phase transition in a coevolving network of conformist and contrarian
voters}
\author{Su Do Yi}
\affiliation{BK21 Physics Research Division and Department of Physics, Sungkyunkwan University, Suwon 440-746, Korea}
\author{Seung Ki Baek}
\affiliation{School of Physics, Korea Institute for Advanced Study, Seoul
130-722, Korea}
\author{Chen-Ping Zhu}
\affiliation{College of Science, Nanjing University of Aeronautics
and Astronautics, Nanjing, 210016, China}
\author{Beom Jun Kim}
\email[Corresponding author, E-mail address: ]{beomjun@skku.edu}
\affiliation{BK21 Physics Research Division and Department of Physics, Sungkyunkwan University, Suwon 440-746, Korea}

\begin{abstract}
In the coevolving voter model, each voter has one of two diametrically
opposite opinions,
and a voter encountering a neighbor with the opposite
opinion may either adopt it or rewire the connection to
another randomly chosen voter sharing the same opinion.
As we smoothly change the relative frequency of rewiring compared to that of
adoption, there occurs a phase transition between an
active phase and a frozen phase. By performing extensive Monte Carlo
calculations, we show that the phase transition is characterized by critical
exponents $\beta = 0.54(1)$ and $\bar\nu = 1.5(1)$, which differ from the
existing mean-field-typed prediction.
We furthermore extend the model by introducing a contrarian type
that tries to have neighbors with the opposite opinion, and
show that the critical behavior still belongs to the same universality class
irrespective of such contrarians' fraction.
\end{abstract}
\pacs{89.75.Fb,05.70.Jk,89.75.Hc,05.65.+b}

\maketitle

\section{INTRODUCTION}
Analyzing social dynamics from a physical viewpoint has become an important
topic in statistical physics. In particular, emergence
of collective opinion is an intriguing phenomenon which can be readily
compared to the collective ordering manifested in many statistical-physical
systems, and the voter model has served as an insightful starting point to
study the opinion dynamics~\cite{*[{See, e.g., }] [{
for further references.}] CastellanoRMP}. When the voter model began to be
investigated in the context of complex networks, the underlying connection
structure among voters was usually assumed to be static. In this case, the
dynamics can be basically described as follows:
The opinion $S_j$ assigned to a voter $j$ is either $+1$ or $-1$ like an
Ising spin. For a given voter $j$, we randomly choose one of the neighbors,
say $k$. If $S_j \neq S_k$, the link between $j$ and $k$ is called active,
and $j$ adopts $k$'s opinion under certain stochastic rules.
As implied in this adoption mechanism,
it is true that our opinion is shaped by whom we are surrounded by, but it
can be equally true that our opinion shapes our social networks. So
the situation becomes more interesting when we allow voters to change
their neighbors. It then looks natural to introduce a tendency to connect
to like-minded people, because as the proverb says, birds of a feather
flock together. This simultaneous change in the network structure is termed
coevolution, and the voter model has been studied in this coevolutionary
framework for the natural reason mentioned
above~\cite{VazquezPRL,DurrettPNAS,*BenczikPRE,*FuPRE,HolmePRE,IniguezPRE}.
Besides the opinion dynamics, we can also mention that the coevolutionary
dynamics plays a crucial role in the context of epidemiology since
adaptation of a network structure in response to disease spreading is
directly related to quarantine policies (see
Refs.~\cite{GrossPRLepi,*GrossROYAL,*CrokIOP,ShawPRE} for theoretical
investigations).
In addition, more physically motivated studies include a coevolving network
of phase oscillators~\cite{AokiPRL}, the kinetic Ising model with Glauber-like
dynamics~\cite{MandraPRE,*HajraEPJB}, and a variation of the $XY$
model~\cite{HolmeYXPRE}.

\section{COEVOLVING NETWORK OF VOTERS}
In this work, we begin with the coevolving voter model proposed by Vazquez \textit{et
al}.~\cite{VazquezPRL} where an active link becomes inert either by rewiring
with probability $p$ or by adoption with probability $1-p$ (see also
Ref.~\cite{ChengCPC,*HerreraEPL,*NardiniPRL,*ZschalerPRE} for its extended
versions). According to a mean-field (MF) argument and numerical
simulations, this model has a well-defined phase transition at a critical
value $p_c$, separating an active phase from a frozen phase in which one
observes no active links between the two different opinions. Specifically,
as $p$ approaches the critical point $p_c$ from below, the active-link
density is expected to scale as $\rho \sim (p_c-p)^{\beta}$, or
equivalently, $\rho \sim N^{-\beta/\bar{\nu}}$ at $p=p_c$ for a large but
finite number $N$ of voters, where $\beta$ and $\bar{\nu}$ are critical
indices. The MF argument in Ref.~\cite{VazquezPRL} predicts
$\beta = 1$ and $p_c=(\mu-2)/(\mu-1)$, where $\mu$ means the average degree.
Although this predicted $p_c$ significantly deviates from numerical
estimates, there are better methods to estimate $p_c$~\cite{BoehmePRE}.
However, it has not been questioned whether the critical behavior is consistent
with the MF prediction, and we will show that this is not the case by
finding $\beta = 0.54(1)$ and $\bar{\nu} = 1.5(1)$. Then, we
extend the model by introducing a peculiar voter type that tries to surround
herself with the opposite opinion, which might be called
heterophily~\cite{KimuraPRE}.
The effect of such behavior has already been studied in opinion
dynamics~\cite{GalamPhysicaA,CrokPHA,*BiswasAX} and Ref.~\cite{LamaEPL} has
argued that such behavior can be generated by stochasticity.
Though this heterophilic dynamics is sometimes dubbed the ``inverse voter''
model~\cite{ZhuPLA}, let us denote the usual voter and the inverse voter
as a conformist ($v$) and a contrarian (${\bar v}$), respectively, following
the terminology in Refs.~\cite{GalamPhysicaA,HongPRL}. Our point is that the
critical behavior is universal regardless of such contrarians' fraction
whereas $p_c$ is a nonuniversal quantity.

\begin{figure}[t]
\includegraphics[width=0.30\textwidth]{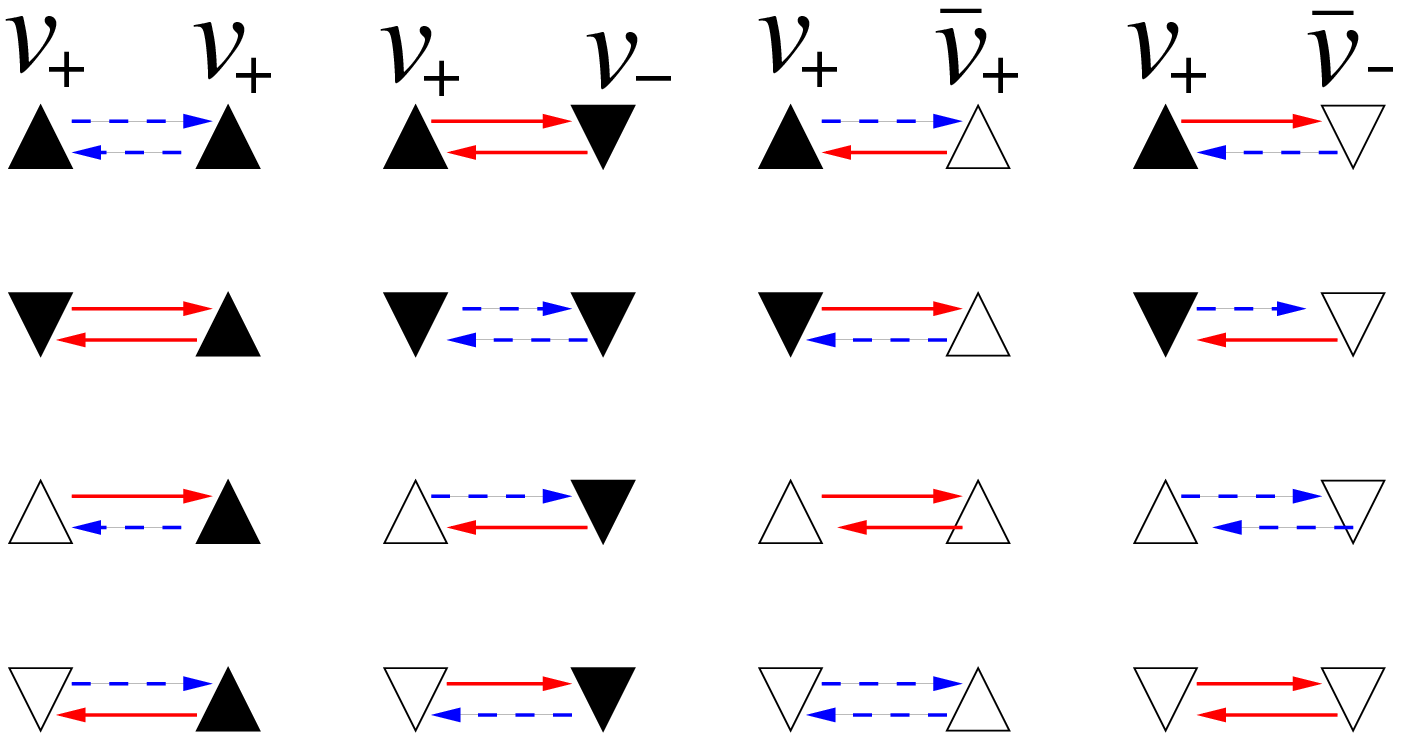}
\caption{(Color online) Possible states of a link between a pair of voters.
We represent a conformist as a filled triangle,
pointing up ($\blacktriangle$) or down ($\blacktriangledown$) depending on
the opinion.
Likewise, a contrarian is represented as an empty triangle
($\vartriangle$ or $\triangledown$).
A full red arrow and a dashed blue arrow
mean an active link and an inert one, respectively.
Whether a link is active depends on whose viewpoint is taken, so
an arrow from a voter means the state of the link from her viewpoint.
For example, the link between $\triangledown$ and
$\blacktriangle$ is active from $\blacktriangle$'s viewpoint, while
it is inert from $\triangledown$'s viewpoint (bottom left).
}
\label{fig:model}
\end{figure}

For simplicity, we assume that voters do not change
their types, (i.e., once a contrarian, forever a contrarian) so
the fraction $r$ of conformists is constant in time. When $r=1$,
our model is reduced to the original one in Ref.~\cite{VazquezPRL}, while
for $r=0$, it is identical to the inverse voter model in Ref.~\cite{ZhuPLA}.
It is important to note that whether a link is active between two voters
depends on their types:
Fig.~\ref{fig:model} shows all the possible link configurations between
two voters, where the signs mean their opinions.
Although each link is expressed as two directed arrows,
there exists only one single link
between a pair of voters so the double arrows imply that the state of a
link is not uniquely defined, in general.
For example, if a conformist with opinion $+1$ is linked to a contrarian
with $-1$, the conformist regards the link as active
while the contrarian regards it as inert.
If a link connecting two voters is regarded by both as active (inert), the link
contributes $1$ ($0$) to the total number of active links in the
system. If only one of them regards it as active, on the other hand, it is
counted as one-half.

Let us now explain details of our calculation.
In terms of a network, a voter is mapped to a node, and a link
connecting nodes $j$ and $k$ can be denoted as $(j,k)$. As an initial
condition, we construct a regular random network of size $N$ where every node
has exactly $\mu$ neighbors (we choose $\mu = 4$ throughout this work).
We then randomly select $rN$ nodes as conformists while the others are set
to be contrarians. Every node has equally random probability to have either
opinion at the starting point. Then, we randomly pick up a node $j$
and choose one of its neighbors, say $k$, also at random.
If the link $(j,k)$ is inert from $j$'s viewpoint, nothing happens.
If it is not, there are two possible options:
With probability $p$, $(j,k)$ is rewired to a randomly
chosen node $l$ in such a way that $(j,l)$ becomes inert from $j$'s viewpoint.
Or with probability $1-p$, $S_j$ is flipped to make $(j,k)$ inert from $j$'s
viewpoint.
At each time step $t$, this updating procedure is repeated for $N$ times.

\section{RESULTS}
We take the active-link density $\rho$ as an order parameter whose
dependence on $N$ at $p=p_c$ provides estimates of the critical indices. In
order to find its stationary value for each $N$, $\rho$ should be first
measured as a function of $t$ over $N_s$ independent samples. Note that we
measure surviving average by including only samples with at least one active
link at each given time $t$.
This surviving-averaged quantity shows transient
behavior at small $t$, whereas it fluctuates violently at large $t$ as
statistics become poor with few surviving samples. So the stationary value
$\rho(N)$ is estimated by taking an average over a certain
time interval $[t_1, t_2]$. In our case, the lower bound $t_1$ is
chosen as $5 \times 10^2$ or $10^3$ depending on $N$, while $t_2$ is fixed
at $5 \times 10^3$. The error bar becomes generally smaller as
$N_s$ increases, and we find it quite enough to work with $N_s = 10^4$ for
$r=1/2$ and $N_s = 2\times 10^5$ for $r=0$ or $r=1$.
In addition,
as for the system size $N$, we need to consider two limitations: That is,
if $N$ is too small, it is difficult to generate many surviving samples at
large $t$ since $\rho = 0$ is an absorbing state which
can always be reached due to fluctuations in a finite-sized system.
If $N$ is too large, on the other hand, it is difficult to approach
a stationary value, especially near the critical point, due to
the diverging relaxation time scale. For these reasons, we will
concentrate on precise estimates of $\rho(N)$ from $N=10^3$ to $N
= 2 \times 10^3$.

\begin{figure}[t]
\includegraphics[width=0.4\textwidth]{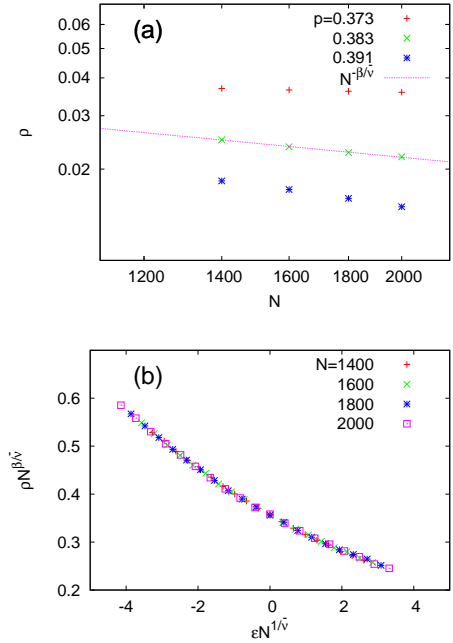}
\caption{(Color online) Coevolving voter model with $r=1$ where all voters
are conformists.
(a) The stationary active-link density $\rho$ is plotted as a function of
$N$ at different rewiring probabilities. The solid line is a guide to eyes.
(b) The function $f$ in Eq.~(\ref{eq:fss}) is obtained by scaling collapse
with $\bar\nu = 1.5(1)$, $\beta = 0.54(1)$, and $p_c = 0.383(1)$.
}
\label{fig:r1}
\end{figure}

\begin{figure}
\includegraphics[width=0.4\textwidth]{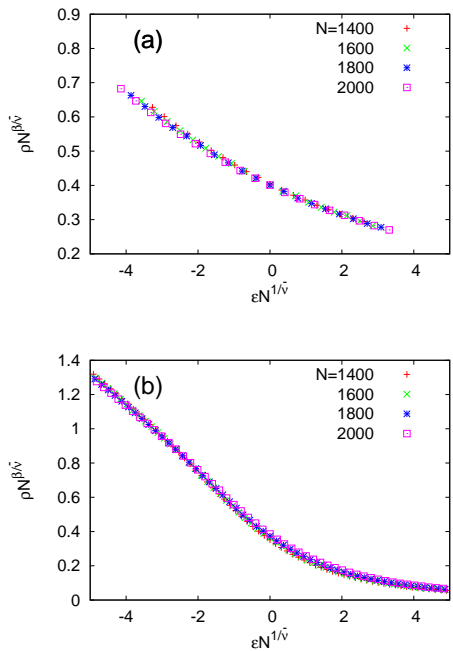}
\caption{(Color online) (a) Scaling collapse for $r=0$ where all
voters are contrarians and (b) for $r=1/2$. The critical
exponents $\beta/\bar\nu = 0.37(2)$ and $\bar\nu = 1.5(1)$ are used in
common, whereas $p_c = 0.383(1)$ in panel (a) in contrast to $p_c =
0.790(1)$ in panel (b).
}
\label{fig:r05}
\end{figure}

\begin{figure}
\includegraphics[width=0.4\textwidth]{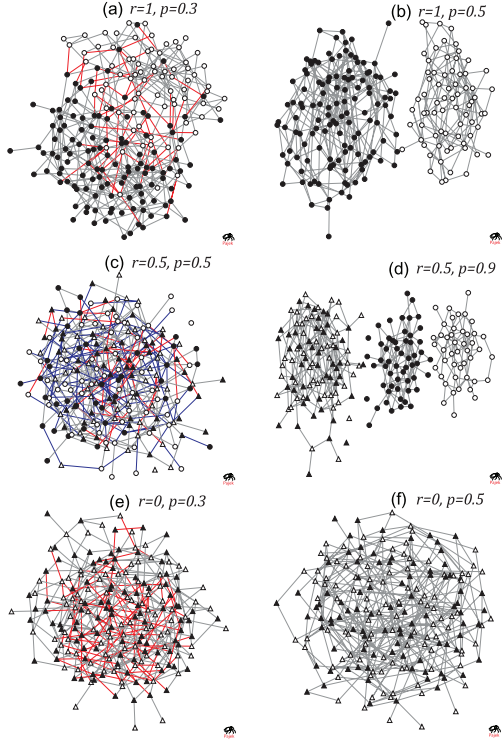}
\caption{(Color online) Snapshots of network structures in the active phase
[(a), (c), (e)] and in the frozen phase [(b), (d), (f)] with $r=1$ [(a),(b)],
$r=1/2$ [(c),(d)], and $r=0$ [(e),(f)]. The number of nodes is $N = 200$ in
each case.
Circles and triangles denote conformists and contrarians, respectively,
and a filled (empty) symbol denotes the opinion $+1$ ($-1$).
Red, blue, and gray colors mean that the link is active in both directions,
active in only one direction, and inert in both directions, respectively.
All links become inert (gray) in the frozen phase.
}
\label{fig:graphs}
\end{figure}

We start with $r=1$ and plot numerical results of $\rho(N)$
at various values of $p$ in Fig.~\ref{fig:r1}(a).
It is observed that $\rho$ decays algebraically at $p \approx
0.383$, indicating the existence of a continuous phase transition at $p_c =
0.383(1)$, in agreement with $p_c \approx 0.38$ found by numerical
simulations in Ref.~\cite{VazquezPRL}.
From the standard finite-size scaling form
\begin{equation}
\rho = N^{-\beta/\bar\nu} f( \varepsilon N^{1/\bar\nu} )
\label{eq:fss}
\end{equation}
with $\varepsilon \equiv (p-p_c)/p_c$,
the slope of the line
in Fig.~\ref{fig:r1}(a) determines $\beta/\bar\nu \approx 0.36$.
We use these values of $p_c$ and $\beta/\bar\nu$ to make the data collapse
in Fig.~\ref{fig:r1}(b). This analysis finds that $p_c = 0.383(1)$,
$\bar\nu = 1.5(1)$, and $\beta = 0.54(1)$.
It is noteworthy that the actual value of $\beta$ is far from the MF prediction
that $\beta = 1$. The reason is that the full mixing
assumption of the MF approach cannot be justified when the system splits
into clusters, which is exactly what happens at the critical point [see
Fig.~\ref{fig:graphs}(a) below]. In this respect, it should not be
surprising that the actual transition is not of the MF type.

We then repeat the same analysis for $r=0$
[Fig.~\ref{fig:r05}(a)]. The resulting
exponents $\beta/\bar\nu = 0.37(2)$ and $\bar\nu = 1.5(1)$ as well as $p_c =
0.383(1)$ are in striking agreement with those in the above case with $r=1$.
From the observation that both with $r=0$ and $r=1$ belong to the
same universality class characterized by the same critical exponents $\beta$
and $\bar\nu$, one might guess that this holds true for any value of $r
\in [0,1]$.
In order to check this, one could perform the same finite-size scaling analysis
for $r=1/2$ [Fig.~\ref{fig:r05}(b)]. Unfortunately, finite-size corrections
to scaling are so substantial in our data that we should be content with a
consistency
check just by tuning $p_c$ with having fixed $\beta/\bar\nu = 0.37$ and
$\bar\nu = 1.5$. In spite of such limitations,
we can say that the result is not inconsistent with the guess.
It is somewhat interesting to note that the coevolving model in
Ref.~\cite{HolmePRE} has $1/\bar\nu \approx 0.7(1)$, which is also close to
our finding $1/\bar\nu \approx 0.67$.

\begin{figure}
\includegraphics[width=0.4\textwidth]{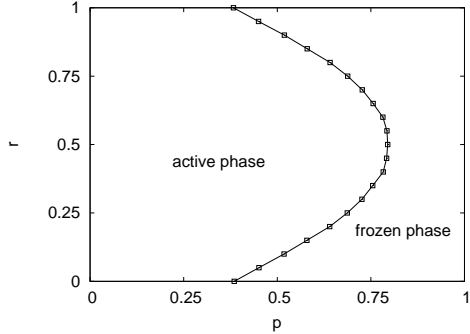}
\caption{Phase diagram of the model on the $(p,r)$ plane. See text
for details.}
\label{fig:phd}
\end{figure}

Let us now turn our attention to determining the critical threshold to
find the shape of the phase boundary.
The estimate of $p_c$ in the simple MF approximation by Vazquez \textit{et
al}.~\cite{VazquezPRL} can be briefly explained in the following way:
Suppose that there is one active link in the system with average degree
$\mu$. For example, there are two clusters of opposite opinions and there is
one single link between them. If the link is rewired with probability $p$,
this active link will disappear. If one voter changes the opinion with
probability $1-p$, then the original active link disappears but all the
other $\mu-1$ links become active. Therefore, the expected change in the
number of active links is $[(\mu-1)-1](1-p) - p$, which is zero below the
threshold, leading to $p_c = (\mu-2)/(\mu-1)$. Note that
the above simple MF approximation does not depend on the value of $r$, and
thus giving $p_c = 2/3$ for $\mu=4$ irrespective of $r$, which is sharply
against our observation that $p_c(r=1/2) \neq p_c(r=1)$.

In order to numerically obtain the actual phase diagram on the $(p,r)$ plane,
we first measure $\rho^\ast \equiv \rho(p = p_c = 0.383)$ at $r=1$ for size
$N=10^3$ and then use the same $\rho^\ast$ to locate $p_c$ for
other values of $r$. In other words, $p_c(r)$ is roughly estimated from the
condition that $\rho(p_c; r, N=10^3) = \rho^\ast$. We have confirmed that
this simple method yields $p_c$ reasonably close to the the above-mentioned
estimates for $r=0$ and $r=1/2$. Furthermore, it is enough to use a smaller
ensemble ($N_s = 200$) for the present purpose. The resulting phase diagram
shows a lobelike shape with a maximum value around $r=1/2$
(Fig.~\ref{fig:phd}), and the MF theory fails to predict this feature once
again.

In order to understand such a shape, let us look into the MF estimate a little
further:
In updating a node with the probability $1-p$, it is possible to come
back to the original configuration by flipping the same node once again. This
possibility is ignored in the simple MF scheme in Ref.~\cite{VazquezPRL},
which means that the spatial correlation among
the newly created active links is ignored.
The $q$-fan motif method~\cite{BoehmePRE} is a
more advanced approximation in the sense that it evolves the configuration
until clusters of the active links are separated by at least one spacing. If
the configuration is expressed with a finite number of separate clusters of
active links after some time, then the method ignores the spatial
correlation among them in describing their further evolution with a
stochastic matrix.
Comparing $r=0$ with $r=1$, we see that the $q$-fan motif method works in
the same way and the resulting estimates of $p_c$ are therefore identical
again. When $0<r<1$, however, there arises an important
difference that there can exist an active link between a conformist and a
contrarian, which never disappears just by changing one's opinion. There is
always a finite chance that a cluster of active links keeps growing from
this active seed link by flipping neighboring voters. Thus the configuration
cannot be generally a collection of finite separated clusters for $0<r<1$,
and the $q$-fan motif method fails in estimating $p_c$. Nevertheless, it
tells us that one needs a higher $p$ in order to suppress such growth
of the cluster by rewiring the active link instead of changing opinions.
This qualitatively explains why the phase boundary at $0<r<1$ is pushed to
higher $p$ in Fig.~\ref{fig:phd}.

Finally, we visualize actual network structures in Fig.~\ref{fig:graphs}.
When everyone is a conformist ($r=1$), one can clearly see that the network
structure radically changes as $p$ passes the critical point so that the
frozen phase is characterized by completely segregated opinions. More
specifically, as $p_c$ is approached from below, two clusters have fewer and
fewer inter-cluster links and they get disconnected at $p_c$. Therefore, in
the vicinity of the critical point, the full-mixing assumption in the simple
MF approximation just breaks down, allowing the critical exponents to
deviate from the MF ones, as confirmed in the present work.
We add a remark that network structures in the frozen phase
largely depend on the conformist fraction $r$. For example, when $r=1/2$,
the system usually splits into three clusters, one for contrarians and the
other two for conformists.

\section{SUMMARY}
In summary, we have numerically studied the phase transition of the
coevolving voter model composed of conformists and contrarians. Our main
result suggests that the simple MF approximation cannot explain the
transition nature, and that the universality class is well characterized by
critical exponents $\beta \approx 0.54$ and $\nu \approx 1.5$, irrespective
of the conformist fraction. We present the full phase diagram on the $(p,r)$
plane, which shows a symmetric lobelike structure around $r=1/2$.

\acknowledgments
B.J.K. was supported by the National Research
Foundation of Korea (NRF) grant funded by the Korea government (MEST) (Grant
No. 2011-0015731) and C.-P.Z. by the Natural Science Foundation
of China (NSFC) (Grant No. 11175086).


\begin{thebibliography}{27}%
\makeatletter
\providecommand \@ifxundefined [1]{%
 \@ifx{#1\undefined}
}%
\providecommand \@ifnum [1]{%
 \ifnum #1\expandafter \@firstoftwo
 \else \expandafter \@secondoftwo
 \fi
}%
\providecommand \@ifx [1]{%
 \ifx #1\expandafter \@firstoftwo
 \else \expandafter \@secondoftwo
 \fi
}%
\providecommand \natexlab [1]{#1}%
\providecommand \enquote  [1]{``#1''}%
\providecommand \bibnamefont  [1]{#1}%
\providecommand \bibfnamefont [1]{#1}%
\providecommand \citenamefont [1]{#1}%
\providecommand \href@noop [0]{\@secondoftwo}%
\providecommand \href [0]{\begingroup \@sanitize@url \@href}%
\providecommand \@href[1]{\@@startlink{#1}\@@href}%
\providecommand \@@href[1]{\endgroup#1\@@endlink}%
\providecommand \@sanitize@url [0]{\catcode `\\12\catcode `\$12\catcode
  `\&12\catcode `\#12\catcode `\^12\catcode `\_12\catcode `\%12\relax}%
\providecommand \@@startlink[1]{}%
\providecommand \@@endlink[0]{}%
\providecommand \url  [0]{\begingroup\@sanitize@url \@url }%
\providecommand \@url [1]{\endgroup\@href {#1}{\urlprefix }}%
\providecommand \urlprefix  [0]{URL }%
\providecommand \Eprint [0]{\href }%
\providecommand \doibase [0]{http://dx.doi.org/}%
\providecommand \selectlanguage [0]{\@gobble}%
\providecommand \bibinfo  [0]{\@secondoftwo}%
\providecommand \bibfield  [0]{\@secondoftwo}%
\providecommand \translation [1]{[#1]}%
\providecommand \BibitemOpen [0]{}%
\providecommand \bibitemStop [0]{}%
\providecommand \bibitemNoStop [0]{.\EOS\space}%
\providecommand \EOS [0]{\spacefactor3000\relax}%
\providecommand \BibitemShut  [1]{\csname bibitem#1\endcsname}%
\let\auto@bib@innerbib\@empty
\bibitem [{\citenamefont {Castellano}\ \emph {et~al.}(2009)\citenamefont
  {Castellano}, \citenamefont {Fortunato},\ and\ \citenamefont
  {Loreto}}]{CastellanoRMP}%
  \BibitemOpen
  \bibfield  {author} {\bibinfo {author} {\bibfnamefont {C.}~\bibnamefont
  {Castellano}}, \bibinfo {author} {\bibfnamefont {S.}~\bibnamefont
  {Fortunato}}, \ and\ \bibinfo {author} {\bibfnamefont {V.}~\bibnamefont
  {Loreto}},\ }\href@noop {} {\bibfield  {journal} {\bibinfo  {journal} {Rev.
  Mod. Phys.}\ }\textbf {\bibinfo {volume} {81}},\ \bibinfo {pages} {591}
  (\bibinfo {year} {2009})}\BibitemShut {NoStop}%
\bibitem [{\citenamefont {Vazquez}\ \emph {et~al.}(2008)\citenamefont
  {Vazquez}, \citenamefont {Egu{\'i}luz},\ and\ \citenamefont
  {San~Miguel}}]{VazquezPRL}%
  \BibitemOpen
  \bibfield  {author} {\bibinfo {author} {\bibfnamefont {F.}~\bibnamefont
  {Vazquez}}, \bibinfo {author} {\bibfnamefont {V.~M.}\ \bibnamefont
  {Egu{\'i}luz}}, \ and\ \bibinfo {author} {\bibfnamefont {M.}~\bibnamefont
  {San~Miguel}},\ }\href@noop {} {\bibfield  {journal} {\bibinfo  {journal}
  {Phys. Rev. Lett.}\ }\textbf {\bibinfo {volume} {100}},\ \bibinfo {pages}
  {108702} (\bibinfo {year} {2008})}\BibitemShut {NoStop}%
\bibitem [{\citenamefont {Durrett}\ \emph {et~al.}(2012)\citenamefont
  {Durrett}, \citenamefont {Gleeson}, \citenamefont {Lloyd}, \citenamefont
  {Mucha}, \citenamefont {Shi}, \citenamefont {Sivakoff}, \citenamefont
  {Socolar},\ and\ \citenamefont {Varghese}}]{DurrettPNAS}%
  \BibitemOpen
  \bibfield  {author} {\bibinfo {author} {\bibfnamefont {R.~R.}\ \bibnamefont
  {Durrett}}, \bibinfo {author} {\bibfnamefont {J.~P.~J.}\ \bibnamefont
  {Gleeson}}, \bibinfo {author} {\bibfnamefont {A.~L.~A.}\ \bibnamefont
  {Lloyd}}, \bibinfo {author} {\bibfnamefont {P.~J.~P.}\ \bibnamefont {Mucha}},
  \bibinfo {author} {\bibfnamefont {F.~F.}\ \bibnamefont {Shi}}, \bibinfo
  {author} {\bibfnamefont {D.~D.}\ \bibnamefont {Sivakoff}}, \bibinfo {author}
  {\bibfnamefont {J.~E. S.~J.}\ \bibnamefont {Socolar}}, \ and\ \bibinfo
  {author} {\bibfnamefont {C.~C.}\ \bibnamefont {Varghese}},\ }\href@noop {}
  {\bibfield  {journal} {\bibinfo  {journal} {Proc. Natl. Acad. Sci. U.S.A.}\ }\textbf
  {\bibinfo {volume} {109}},\ \bibinfo {pages} {3682} (\bibinfo {year}
  {2012})}\BibitemShut {NoStop}%
\bibitem [{\citenamefont {Benczik}\ \emph {et~al.}(2009)\citenamefont
  {Benczik}, \citenamefont {Benczik}, \citenamefont {Schmittmann},\ and\
  \citenamefont {Zia}}]{BenczikPRE}%
  \BibitemOpen
  \bibfield  {author} {\bibinfo {author} {\bibfnamefont {I.}~\bibnamefont
  {Benczik}}, \bibinfo {author} {\bibfnamefont {S.}~\bibnamefont {Benczik}},
  \bibinfo {author} {\bibfnamefont {B.}~\bibnamefont {Schmittmann}}, \ and\
  \bibinfo {author} {\bibfnamefont {R.}~\bibnamefont {Zia}},\ }\href@noop {}
  {\bibfield  {journal} {\bibinfo  {journal} {Phys. Rev. E}\ }\textbf {\bibinfo
  {volume} {79}},\ \bibinfo {pages} {046104} (\bibinfo {year}
  {2009})}\BibitemShut {NoStop}%
\bibitem [{\citenamefont {Fu}\ and\ \citenamefont {Wang}(2008)}]{FuPRE}%
  \BibitemOpen
  \bibfield  {author} {\bibinfo {author} {\bibfnamefont {F.}~\bibnamefont
  {Fu}}\ and\ \bibinfo {author} {\bibfnamefont {L.}~\bibnamefont {Wang}},\
  }\href@noop {} {\bibfield  {journal} {\bibinfo  {journal} {Phys. Rev. E}\
  }\textbf {\bibinfo {volume} {78}},\ \bibinfo {pages} {016104} (\bibinfo
  {year} {2008})}\BibitemShut {NoStop}%
\bibitem [{\citenamefont {Holme}\ and\ \citenamefont
  {Newman}(2006)}]{HolmePRE}%
  \BibitemOpen
  \bibfield  {author} {\bibinfo {author} {\bibfnamefont {P.}~\bibnamefont
  {Holme}}\ and\ \bibinfo {author} {\bibfnamefont {M.~E.~J.}\ \bibnamefont
  {Newman}},\ }\href@noop {} {\bibfield  {journal} {\bibinfo  {journal} {Phys.
  Rev. E}\ }\textbf {\bibinfo {volume} {74}},\ \bibinfo {pages} {056108}
  (\bibinfo {year} {2006})}\BibitemShut {NoStop}%
\bibitem [{\citenamefont {I{\~n}iguez}\ \emph {et~al.}(2011)\citenamefont
  {I{\~n}iguez}, \citenamefont {Kert{\'e}sz}, \citenamefont {Kaski},\ and\
  \citenamefont {Barrio}}]{IniguezPRE}%
  \BibitemOpen
  \bibfield  {author} {\bibinfo {author} {\bibfnamefont {G.}~\bibnamefont
  {I{\~n}iguez}}, \bibinfo {author} {\bibfnamefont {J.}~\bibnamefont
  {Kert{\'e}sz}}, \bibinfo {author} {\bibfnamefont {K.}~\bibnamefont {Kaski}},
  \ and\ \bibinfo {author} {\bibfnamefont {R.}~\bibnamefont {Barrio}},\
  }\href@noop {} {\bibfield  {journal} {\bibinfo  {journal} {Phys. Rev. E}\
  }\textbf {\bibinfo {volume} {83}},\ \bibinfo {pages} {016111} (\bibinfo
  {year} {2011})}\BibitemShut {NoStop}%
\bibitem [{\citenamefont {Gross}\ \emph {et~al.}(2006)\citenamefont {Gross},
  \citenamefont {D'Lima},\ and\ \citenamefont {Blasius}}]{GrossPRLepi}%
  \BibitemOpen
  \bibfield  {author} {\bibinfo {author} {\bibfnamefont {T.}~\bibnamefont
  {Gross}}, \bibinfo {author} {\bibfnamefont {C.~J.~D.}\ \bibnamefont
  {D'Lima}}, \ and\ \bibinfo {author} {\bibfnamefont {B.}~\bibnamefont
  {Blasius}},\ }\href {\doibase 10.1103/PhysRevLett.96.208701} {\bibfield
  {journal} {\bibinfo  {journal} {Phys. Rev. Lett.}\ }\textbf {\bibinfo
  {volume} {96}},\ \bibinfo {pages} {208701} (\bibinfo {year}
  {2006})}\BibitemShut {NoStop}%
\bibitem [{\citenamefont {Gross}\ and\ \citenamefont
  {Blasius}(2008)}]{GrossROYAL}%
  \BibitemOpen
  \bibfield  {author} {\bibinfo {author} {\bibfnamefont {T.}~\bibnamefont
  {Gross}}\ and\ \bibinfo {author} {\bibfnamefont {B.}~\bibnamefont
  {Blasius}},\ }\href {\doibase 10.1098/rsif.2007.1229} {\bibfield  {journal}
  {\bibinfo  {journal} {Journal of The Royal Society Interface}\ }\textbf
  {\bibinfo {volume} {5}},\ \bibinfo {pages} {259} (\bibinfo {year}
  {2008})}\BibitemShut {NoStop}%
\bibitem [{\citenamefont {Crokidakis}\ and\ \citenamefont
  {Queirós}(2012)}]{CrokIOP}%
  \BibitemOpen
  \bibfield  {author} {\bibinfo {author} {\bibfnamefont {N.}~\bibnamefont
  {Crokidakis}}\ and\ \bibinfo {author} {\bibfnamefont {S.~M.~D.}\ \bibnamefont
  {Queirós}},\ }\href {http://stacks.iop.org/1742-5468/2012/i=06/a=P06003}
  {\bibfield  {journal} {\bibinfo  {journal} {J. Stat. Mech: Theory Exp.}\ }\textbf
  {\bibinfo {volume} {2012}},\ \bibinfo {pages} {P06003} (\bibinfo {year}
  {2012})}\BibitemShut {NoStop}%
\bibitem [{\citenamefont {Shaw}\ and\ \citenamefont
  {Schwartz}(2008)}]{ShawPRE}%
  \BibitemOpen
  \bibfield  {author} {\bibinfo {author} {\bibfnamefont {L.~B.}\ \bibnamefont
  {Shaw}}\ and\ \bibinfo {author} {\bibfnamefont {I.~B.}\ \bibnamefont
  {Schwartz}},\ }\href {\doibase 10.1103/PhysRevE.77.066101} {\bibfield
  {journal} {\bibinfo  {journal} {Phys. Rev. E}\ }\textbf {\bibinfo {volume}
  {77}},\ \bibinfo {pages} {066101} (\bibinfo {year} {2008})}\BibitemShut
  {NoStop}%
\bibitem [{\citenamefont {Aoki}\ and\ \citenamefont {Aoyagi}(2009)}]{AokiPRL}%
  \BibitemOpen
  \bibfield  {author} {\bibinfo {author} {\bibfnamefont {T.}~\bibnamefont
  {Aoki}}\ and\ \bibinfo {author} {\bibfnamefont {T.}~\bibnamefont {Aoyagi}},\
  }\href@noop {} {\bibfield  {journal} {\bibinfo  {journal} {Phys. Rev. Lett.}\
  }\textbf {\bibinfo {volume} {102}},\ \bibinfo {pages} {034101} (\bibinfo
  {year} {2009})}\BibitemShut {NoStop}%
\bibitem [{\citenamefont {Mandr\`a}\ \emph {et~al.}(2009)\citenamefont
  {Mandr\`a}, \citenamefont {Fortunato},\ and\ \citenamefont
  {Castellano}}]{MandraPRE}%
  \BibitemOpen
  \bibfield  {author} {\bibinfo {author} {\bibfnamefont {S.}~\bibnamefont
  {Mandr\`a}}, \bibinfo {author} {\bibfnamefont {S.}~\bibnamefont {Fortunato}},
  \ and\ \bibinfo {author} {\bibfnamefont {C.}~\bibnamefont {Castellano}},\
  }\href {\doibase 10.1103/PhysRevE.80.056105} {\bibfield  {journal} {\bibinfo
  {journal} {Phys. Rev. E}\ }\textbf {\bibinfo {volume} {80}},\ \bibinfo
  {pages} {056105} (\bibinfo {year} {2009})}\BibitemShut {NoStop}%
\bibitem [{\citenamefont {Hajra}\ and\ \citenamefont
  {Chandra}(2012)}]{HajraEPJB}%
  \BibitemOpen
  \bibfield  {author} {\bibinfo {author} {\bibfnamefont {K.~B.}\ \bibnamefont
  {Hajra}}\ and\ \bibinfo {author} {\bibfnamefont {A.~K.}\ \bibnamefont
  {Chandra}},\ }\href@noop {} {\bibfield  {journal} {\bibinfo  {journal} {Eur.
  Phys. J. B}\ }\textbf {\bibinfo {volume} {85}},\ \bibinfo {pages} {27}
  (\bibinfo {year} {2012})}\BibitemShut {NoStop}%
\bibitem [{\citenamefont {Holme}\ \emph {et~al.}(2009)\citenamefont {Holme},
  \citenamefont {Wu},\ and\ \citenamefont {Minnhagen}}]{HolmeYXPRE}%
  \BibitemOpen
  \bibfield  {author} {\bibinfo {author} {\bibfnamefont {P.}~\bibnamefont
  {Holme}}, \bibinfo {author} {\bibfnamefont {Z.-X.}\ \bibnamefont {Wu}}, \
  and\ \bibinfo {author} {\bibfnamefont {P.}~\bibnamefont {Minnhagen}},\ }\href
  {\doibase 10.1103/PhysRevE.80.036120} {\bibfield  {journal} {\bibinfo
  {journal} {Phys. Rev. E}\ }\textbf {\bibinfo {volume} {80}},\ \bibinfo
  {pages} {036120} (\bibinfo {year} {2009})}\BibitemShut {NoStop}%
\bibitem [{\citenamefont {Cheng}\ \emph {et~al.}(2010)\citenamefont {Cheng},
  \citenamefont {Hu}, \citenamefont {Di},\ and\ \citenamefont
  {Fan}}]{ChengCPC}%
  \BibitemOpen
  \bibfield  {author} {\bibinfo {author} {\bibfnamefont {J.}~\bibnamefont
  {Cheng}}, \bibinfo {author} {\bibfnamefont {Y.}~\bibnamefont {Hu}}, \bibinfo
  {author} {\bibfnamefont {Z.}~\bibnamefont {Di}}, \ and\ \bibinfo {author}
  {\bibfnamefont {Y.}~\bibnamefont {Fan}},\ }\href@noop {} {\bibfield
  {journal} {\bibinfo  {journal} {Comp. Phys. Comm.}\ }\textbf {\bibinfo
  {volume} {181}},\ \bibinfo {pages} {1697} (\bibinfo {year}
  {2010})}\BibitemShut {NoStop}%
\bibitem [{\citenamefont {Herrera}\ \emph {et~al.}(2011)\citenamefont
  {Herrera}, \citenamefont {Cosenza}, \citenamefont {Tucci},\ and\
  \citenamefont {Gonz{\'a}lez-Avella}}]{HerreraEPL}%
  \BibitemOpen
  \bibfield  {author} {\bibinfo {author} {\bibfnamefont {J.~L.}\ \bibnamefont
  {Herrera}}, \bibinfo {author} {\bibfnamefont {M.~G.}\ \bibnamefont
  {Cosenza}}, \bibinfo {author} {\bibfnamefont {K.}~\bibnamefont {Tucci}}, \
  and\ \bibinfo {author} {\bibfnamefont {J.~C.}\ \bibnamefont
  {Gonz{\'a}lez-Avella}},\ }\href@noop {} {\bibfield  {journal} {\bibinfo
  {journal} {EPL}\ }\textbf {\bibinfo {volume} {95}},\ \bibinfo {pages} {58006}
  (\bibinfo {year} {2011})}\BibitemShut {NoStop}%
\bibitem [{\citenamefont {Nardini}\ \emph {et~al.}(2008)\citenamefont
  {Nardini}, \citenamefont {Kozma},\ and\ \citenamefont {Barrat}}]{NardiniPRL}%
  \BibitemOpen
  \bibfield  {author} {\bibinfo {author} {\bibfnamefont {C.}~\bibnamefont
  {Nardini}}, \bibinfo {author} {\bibfnamefont {B.}~\bibnamefont {Kozma}}, \
  and\ \bibinfo {author} {\bibfnamefont {A.}~\bibnamefont {Barrat}},\
  }\href@noop {} {\bibfield  {journal} {\bibinfo  {journal} {Phys. Rev. Lett.}\
  }\textbf {\bibinfo {volume} {100}},\ \bibinfo {pages} {158701} (\bibinfo
  {year} {2008})}\BibitemShut {NoStop}%
\bibitem [{\citenamefont {Zschaler}\ \emph {et~al.}(2012)\citenamefont
  {Zschaler}, \citenamefont {B{\"o}hme}, \citenamefont {Sei{\ss}inger},
  \citenamefont {Huepe},\ and\ \citenamefont {Gross}}]{ZschalerPRE}%
  \BibitemOpen
  \bibfield  {author} {\bibinfo {author} {\bibfnamefont {G.}~\bibnamefont
  {Zschaler}}, \bibinfo {author} {\bibfnamefont {G.~A.}\ \bibnamefont
  {B{\"o}hme}}, \bibinfo {author} {\bibfnamefont {M.}~\bibnamefont
  {Sei{\ss}inger}}, \bibinfo {author} {\bibfnamefont {C.}~\bibnamefont
  {Huepe}}, \ and\ \bibinfo {author} {\bibfnamefont {T.}~\bibnamefont
  {Gross}},\ }\href@noop {} {\bibfield  {journal} {\bibinfo  {journal} {Phys.
  Rev. E}\ }\textbf {\bibinfo {volume} {85}},\ \bibinfo {pages} {046107}
  (\bibinfo {year} {2012})}\BibitemShut {NoStop}%
\bibitem [{\citenamefont {B{\"o}hme}\ and\ \citenamefont
  {Gross}(2011)}]{BoehmePRE}%
  \BibitemOpen
  \bibfield  {author} {\bibinfo {author} {\bibfnamefont {G.~A.}\ \bibnamefont
  {B{\"o}hme}}\ and\ \bibinfo {author} {\bibfnamefont {T.}~\bibnamefont
  {Gross}},\ }\href@noop {} {\bibfield  {journal} {\bibinfo  {journal} {Phys.
  Rev. E}\ }\textbf {\bibinfo {volume} {83}},\ \bibinfo {pages} {035101}
  (\bibinfo {year} {2011})}\BibitemShut {NoStop}%
\bibitem [{\citenamefont {Kimura}\ and\ \citenamefont
  {Hayakawa}(2008)}]{KimuraPRE}%
  \BibitemOpen
  \bibfield  {author} {\bibinfo {author} {\bibfnamefont {D.}~\bibnamefont
  {Kimura}}\ and\ \bibinfo {author} {\bibfnamefont {Y.}~\bibnamefont
  {Hayakawa}},\ }\href@noop {} {\bibfield  {journal} {\bibinfo  {journal}
  {Phys. Rev. E}\ }\textbf {\bibinfo {volume} {78}},\ \bibinfo {pages} {016103}
  (\bibinfo {year} {2008})}\BibitemShut {NoStop}%
\bibitem [{\citenamefont {Galam}(2004)}]{GalamPhysicaA}%
  \BibitemOpen
  \bibfield  {author} {\bibinfo {author} {\bibfnamefont {S.}~\bibnamefont
  {Galam}},\ }\href@noop {} {\bibfield  {journal} {\bibinfo  {journal} {Physica
  A}\ }\textbf {\bibinfo {volume} {333}},\ \bibinfo {pages} {453} (\bibinfo
  {year} {2004})}\BibitemShut {NoStop}%
\bibitem [{\citenamefont {Crokidakis}\ and\ \citenamefont
  {Anteneodo}(2012)}]{CrokPHA}%
  \BibitemOpen
  \bibfield  {author} {\bibinfo {author} {\bibfnamefont {N.}~\bibnamefont
  {Crokidakis}}\ and\ \bibinfo {author} {\bibfnamefont {C.}~\bibnamefont
  {Anteneodo}},\ } \href{\doibase 10.1103/PhysRevE.86.061127} 
  {\bibfield  {journal} {\bibinfo  {journal} {Phys. Rev. E}\ } 
  \textbf {\bibinfo {volume} {86}},\ \bibinfo
  {pages} {061127} (\bibinfo {year} {2012})}\BibitemShut {NoStop}%
\bibitem [{\citenamefont {Biswas}\ \emph {et~al.}(2012)\citenamefont {Biswas},
  \citenamefont {Chatterjee},\ and\ \citenamefont {Sen}}]{BiswasAX}%
  \BibitemOpen
  \bibfield  {author} {\bibinfo {author} {\bibfnamefont {S.}~\bibnamefont
  {Biswas}}, \bibinfo {author} {\bibfnamefont {A.}~\bibnamefont {Chatterjee}},
  \ and\ \bibinfo {author} {\bibfnamefont {P.}~\bibnamefont {Sen}},\ }\href
  {\doibase 10.1016/j.physa.2012.01.046} {\bibfield  {journal} {\bibinfo
  {journal} {Physica A}\ }\textbf {\bibinfo {volume} {391}},\ \bibinfo {pages}
  {3257 } (\bibinfo {year} {2012})}\BibitemShut {NoStop}%
\bibitem [{\citenamefont {De~La~Lama}\ \emph {et~al.}(2007)\citenamefont
  {De~La~Lama}, \citenamefont {L{\'o}pez},\ and\ \citenamefont
  {Wio}}]{LamaEPL}%
  \BibitemOpen
  \bibfield  {author} {\bibinfo {author} {\bibfnamefont {M.~S.}\ \bibnamefont
  {De~La~Lama}}, \bibinfo {author} {\bibfnamefont {J.~M.}\ \bibnamefont
  {L{\'o}pez}}, \ and\ \bibinfo {author} {\bibfnamefont {H.~S.}\ \bibnamefont
  {Wio}},\ }\href@noop {} {\bibfield  {journal} {\bibinfo  {journal} {EPL}\
  }\textbf {\bibinfo {volume} {72}},\ \bibinfo {pages} {851} (\bibinfo {year}
  {2007})}\BibitemShut {NoStop}%
\bibitem [{\citenamefont {Zhu}\ \emph {et~al.}(2011)\citenamefont {Zhu},
  \citenamefont {Kong}, \citenamefont {Li}, \citenamefont {Gu},\ and\
  \citenamefont {Xiong}}]{ZhuPLA}%
  \BibitemOpen
  \bibfield  {author} {\bibinfo {author} {\bibfnamefont {C.-P.}\ \bibnamefont
  {Zhu}}, \bibinfo {author} {\bibfnamefont {H.}~\bibnamefont {Kong}}, \bibinfo
  {author} {\bibfnamefont {L.}~\bibnamefont {Li}}, \bibinfo {author}
  {\bibfnamefont {Z.-M.}\ \bibnamefont {Gu}}, \ and\ \bibinfo {author}
  {\bibfnamefont {S.-J.}\ \bibnamefont {Xiong}},\ }\href@noop {} {\bibfield
  {journal} {\bibinfo  {journal} {Phys. Lett.}\ }\textbf {\bibinfo {volume}
  {A375}},\ \bibinfo {pages} {1378} (\bibinfo {year} {2011})}\BibitemShut
  {NoStop}%
\bibitem [{\citenamefont {Hong}\ and\ \citenamefont
  {Strogatz}(2011)}]{HongPRL}%
  \BibitemOpen
  \bibfield  {author} {\bibinfo {author} {\bibfnamefont {H.}~\bibnamefont
  {Hong}}\ and\ \bibinfo {author} {\bibfnamefont {S.}~\bibnamefont
  {Strogatz}},\ }\href@noop {} {\bibfield  {journal} {\bibinfo  {journal}
  {Phys. Rev. Lett.}\ }\textbf {\bibinfo {volume} {106}},\ \bibinfo {pages}
  {054102} (\bibinfo {year} {2011})}\BibitemShut {NoStop}%
\end{thebibliography}
%

\end{document}